\documentstyle[preprint,aps,floats]{revtex}
\tightenlines
\input{epsf.tex}
\begin{document}
%\rightline{DAMTP-1999-50}
\def\ba{\begin{eqnarray}}
\def\ea{\end{eqnarray}}
\def\be{\begin{equation}}
\def\ee{\end{equation}}
\def\tr{{\rm tr}}
\def\sech{{\rm sech }}
\def\gtorder{\mathrel{\raise.3ex\hbox{$>$}\mkern-14mu
             \lower0.6ex\hbox{$\sim$}}}
\def\ltorder{\mathrel{\raise.3ex\hbox{$<$}\mkern-14mu
             \lower0.6ex\hbox{$\sim$}}}
\title{Cosmological Perturbations Generated in the 
Colliding Bubble Braneworld Universe}

\author{Jose J. Blanco-Pillado\thanks{E-mail:
J.J.Blanco-Pillado@damtp.cam.ac.uk}
and Martin Bucher\thanks{E-mail: M.A.Bucher@damtp.cam.ac.uk}\\
DAMTP, Centre for Mathematical Sciences, University of Cambridge\\
Wilberforce Road, Cambridge CB3 0WA, United Kingdom}

\date{12 November 2001; Revised 18 December 2001}

\maketitle

\begin{abstract}%
We compute the cosmological perturbations generated in the colliding bubble
braneworld universe in which bubbles filled with five-dimensional
anti-de Sitter space $(AdS^5)$ expanding within a
five dimensional de Sitter space $(dS^5)$
or Minkowski space $(M^5)$ collide to form a (3+1) dimensional {\it local brane}
on which the cosmology is virtually identical to that of the Randall-Sundrum model.
The perturbation calculation presented here is valid to linear order
but treats the fluctuations of the expanding bubbles as (3+1) dimensional
fields localized on the bubble wall. 
We find that for bubbles expanding in $dS^5$ the dominant contribution to 
the power spectrum is `red', with $dP(k)/d[log(k)]\approx 1/(m_4\ell )^2
[1+(R/\ell _{ext})^2]^2(\ell /R)^2(k_c/k)^2$
where $R$ is the spatial curvature radius of the 
universe on the local brane at the moment of bubble collision, 
$\ell $ is the curvature radius
of the bulk $AdS^5$ space within the bubbles, $\ell _{ext}$ is the curvature
radius of the $dS^5$ (taken as infinite for the $M^5$ case), $k_c$
is the wavenumber of the spatial curvature scale, and $m_4$
is the four-dimensional Planck mass. The perturbations are minuscule and 
well below observational detection except when $(m_4\ell )$ is not
large or in certain cases where $R$ 
at the moment of collision exceeds $\bar R=\ell _{ext}(\ell _{ext}/\ell ).$
{\bf Note:} This paper supersedes a previous version titled ``Exactly Scale-Invariant
Cosmological Perturbations From a Colliding Bubble Braneworld Universe'' in which we
erroneously claimed that a scale-invariant spectrum results for the case of bubbles
expanding in $M^5.$ This paper corrects the errors of the previous version
and extends the analysis to the more interesting and general case of bubbles 
expanding in $dS^5.$
\end{abstract}

\vskip 0.5in 

\section{Introduction}

One of us (MB) recently proposed\cite{Bucher} that the collision 
of two bubbles of a true anti-de Sitter space vacuum 
expanding within either de Sitter or Minkowski space can
give rise to a braneworld universe consisting of a $(3+1)$-dimensional 
FLRW brane universe embedded in $(4+1)$-dimensional anti-de Sitter 
space. A $Z_2$ (or some larger) discrete symmetry unbroken outside the bubbles
but broken in the AdS phase inside requires that after the collision a brane or 
domain wall form between the two bubbles when the choice of vacua
of the colliding bubbles differs. This is the (3+1)-dimensional brane universe on 
which we live. The spacetime contained within the future 
lightcone of an observer situated on the $(3+1)$-dimensional brane universe
resulting after the brane collision is virtually identical to that of
Randall-Sundrum (RS)\cite{Randall} cosmogony\cite{langloisa}. 
Consequently, gravity on
the brane\cite{Garriga-Tanaka} and the results of any experiment
performed there will be identical to those in the infinite, one-brane
RS scenario. In the present scenario, however, the causal past of the local 
brane, and in particular how certain preferred initial conditions are 
established, is completely distinct.

The standard RS cosmology suffers from the bulk smoothness and horizon 
problems\cite{trodden}. Although inflation on the $(3+1)$-dimensional brane
proposed in some braneworld cosmologies can smooth out whatever inhomogeneities
may have previously existed on the brane, it remains a complete puzzle 
why the $(4+1)$-dimensional bulk into which this brane expands should initially be homogeneous and
isotropic. If the bulk is not homogeneous and isotropic at the outset,
inhomogeneities in the bulk induce inhomogeneities on the brane
at late times. The usual Randall-Sundrum scenario is also plagued with timelike
boundaries at infinity and it is unclear what sort of boundary
conditions should be imposed on these \cite{hawking-ellis}. 
In the colliding bubble scenario
these problems are avoided because a special initial state is 
singled out. In the de Sitter case, the five-dimensional 
Bunch-Davies (BD)\cite{bunch-davies}
vacuum is singled out because this state is an attractor.
For Minkowski space, there is a unique, well-defined vacuum
to be used as an initial state.
Consequently, our proposal resolves the five-dimensional bulk smoothness 
and horizon
problems. A braneworld arising within the framework of the ``no boundary"
proposal suggests another possible
way to get around this problem.\cite{gtnb,hertog}

In ref.\cite{Bucher} the colliding bubble braneworld universe was presented and 
a heuristic discussion of the physical mechanisms 
underlying the generation of cosmological 
density perturbations was given. The colliding bubble 
braneworld universe heavily
relies of the dynamics of false vacuum decay, elucidated quite some time ago
by a number of authors\cite{fvd}. The idea of a cosmology in which entropy and 
matter-radiation is generated from brane collisions was proposed in the work of
Dvali and Tye\cite{dvali} on brane inflation and in the work of 
Khoury, Ovrut, Steinhardt, and Turok\cite{turok} on the ekpyrotic
scenario. Perkins\cite{perkins} considered a braneworld on an expanding bubble;
however, unlike here, in his scenario bubble collisions are regarded as
cataclysmic events to be avoided.  Gorsky and Selivanov\cite{gorsky} 
considered some similar ideas 
involving a uniform external field and the Schwinger mechanism 
and Euclidean instantons.
More recently, Gen et al.\cite{tanaka} have proposed a interesting
variation on the colliding bubble braneworld scenario in which two bubbles 
collide, one nucleating within the other.

\begin{figure}
\begin{center}
\epsfxsize=3in
\epsfysize=3in
\leavevmode\epsfbox{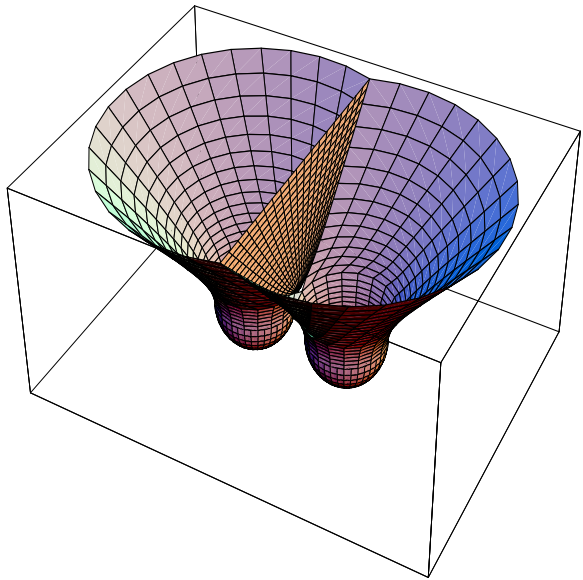}
\end{center}
\caption{{\bf Bubble Collision Geometry.}
This figure shows a $(2+1)$-dimensional spacetime diagram
illustrating the braneworld cosmology contemplated in this paper.
Two spatial dimensions have been suppressed.}
\label{fig:0}
\end{figure}

In this article we present a quantitative calculation of cosmological density 
perturbations generated in this model subject to two simplifying assumptions.
Firstly, we assume that the perturbations are linear. This assumption
is well justified. Secondly, we ignore the self-gravitation of the 
perturbations of the expanding bubbles, although gravity is
completely taken into account for the zeroth order, unperturbed 
solution. As we shall see, ignoring the self-gravity of the perturbations
is rather a questionable approximation. 
In all cases of interest, dimensional arguments suggest
that gravity may play an important role in the dynamics of the 
expanding bubble perturbations at linear order. 
It is not clear whether such corrections 
decouple for some miraculous reason, whether they 
merely result in an $O(1)$ renormalization of the overall
amplitude, or whether they spoil some of the apparently 
delicate cancellations present in the calculation, generating 
perturbations of differing magnitude and spectral character. 
Only by carrying out a fully five-dimensional calculation of the density 
perturbations, taking into account the couplings of each mode
of the bubble wall to an infinite number of modes of 
five-dimensional gravity of the same total angular momentum,
will it be possible to answer this question definitively.
Nevertheless, given the technical difficulty of such a calculation, we
still consider it worthwhile to calculate the linear perturbations 
within the framework of a (3+1) dimensional theory. Interestingly,
the perturbations thus calculated are small, in most of 
parameter space well below the magnitude of those observed in our
universe. Their spectral character, however, is of the wrong type, suggesting
the necessity of some other mechanism to generate the known
cosmological perturbations.

The organization of this paper is as follows. Section II 
describes how perturbations prior to the bubble collision may
be described as scalar fields on the two unperturbed expanding 
bubbles and how these fluctuations of the bubble
walls translate into density perturbations
at the moment of bubble collision. We also note the possible importance
of gravitational corrections to this picture. In section III we
show how to describe the Bunch-Davies vacuum for this scalar
field in terms of the appropriate mode expansion. In section IV
these results are combined to obtain the scalar power spectrum
for bubbles expanding in five-dimensional Minkowski space $(M^5).$ 
Section V generalizes to the case of bubbles expanding in 
five-dimensional de Sitter space $(dS^5).$ Finally, in section
VI we present some concluding remarks. 

\section{Bubble Wall Vacuum Fluctuations and Junction Conditions 
at the Collision}

We begin by discussing the perturbations on the 
expanding bubbles before collision. As explained
in the work of Garriga and Vilenkin\cite{garrigaa,garrigaab}, if one ignores
the coupling to gravity, there is but a single
degree of freedom, that of translations of the bubble
wall, completely described to the linear order of 
interest here by a free scalar field that lives on 
the bubble wall. The bubble wall, idealized here as
infinitely thin, traces out a world volume trajectory
of the shape of a hyperboloid (i.e., the locus of all points 
at a certain proper geodesic distance from the nucleation
center). In the absence of perturbations,  this hyperboloid is endowed with 
the internal geometry of de Sitter space. Perturbations
are described by means of a scalar field $\chi $ defined
on the unperturbed hyperboloid ${\cal{H}}.$ At each point $p$ on ${\cal{H}},$
$\chi $ is assigned a value equal to the distance to the
perturbed brane worldvolume along the geodesic passing 
through $p$ normal to ${\cal{H}},$ with $\chi $ taken positive
for outward displacements.
Of course, since each bubble nucleates
as the result of quantum tunnelling, this hyperboloid
does not really extend infinitely far into the past. 
The bubble materializes through quantum tunnelling
within a compact (bounded) region. But it is not possible 
to determine in which rest frame the bubble has nucleated
without seriously disturbing the outcome of the quantum
tunnelling process. Consequently, the quantum state
of the fluctuations must be $SO(4,1)$ invariant.
This requirement uniquely fixes the quantum state of
the fluctuations.  It follows that the scalar field $\chi $ is in the 
Bunch-Davies vacuum. The bubble wall tension $\tau $ appears in
 the action for $\chi $ as an overall linear factor with units
${\rm (mass)}^4,$ so that $\phi =\tau ^{1/2}\chi $ is the 
customarily normalized $(3+1)$-dimensional scalar field
with units of ${\rm (energy)}^1.$ As pointed out by Garriga and 
Vilenkin, the mass of this scalar field is $m^2 = - 4H_{b}^2$ (where
$H_b$ is the Hubble constant on the bubble surface) is completely fixed
by symmetry. Since the $l=1$ modes correspond
to translation of the bubble, the change in action for these modes
must vanish. This requirement completely fixes the mass.

Treating the perturbations of the expanding bubble wall as a scalar
field in the manner just described would be exact if the 
fluctuations of the bubble did not gravitate. As pointed out 
in ref.~\cite{garrigaab} and demonstrated explicitly in 
appendix A of this paper, the relation $m^2 = - 4H_{b}^2$ 
continues to hold for weakly gravitating bubbles in 
$AdS^5$ and $dS^5,$ and more generally remains exact
for the $l=0$ and $l=1$ solutions corresponding 
to rigid spacetime translations of the nucleation centers
of the bubble, no matter how strong the self-gravity. 
However, when the bubble self-gravity 
is taken into account, leading to a jump in the extrinsic
curvature across the two sides as indicated by the Israel
matching condition, this description ceases to be exact 
for the remaining modes. The inconsistency of the above solution
in the presence of self-gravity of the wall can be seen
by comparing the change in area $\delta A/A$ as computed
on the two sides having differing extrinsic curvature.
These two quantities must be the same but are not.
Physically, for the $l=2$ and higher modes,
as the bubble wall oscillates, gravity waves are emitted.
To obtain a simple estimate to what extent it is justified to 
ignore self-gravity, we compare the order of magnitude of the 
Newtonian gravitational self-energy of the critical bubble $E_{G}$
to its characteristic non-gravitational energy 
$E_{NG}.$ Let $\ell $ be the curvature radius of
the $AdS^5.$ The tension of the bubble wall is of the same
order as the cosmological constant required on the local 
brane so that at late times its geometry is that of $M^4,$
that is $\tau \approx \Lambda _4={m_5}^3/\ell ,$ where $m_5$ is the 
five dimensional Planck mass, related to the 
four-dimensional Planck mass $m_4$ through 
${m_4}^2={m_5}^3\ell .$ It cannot be smaller, for otherwise
the local brane would be unstable against bifurcations into two
bubble wall branes.
It follows that the critical bubble radius is of
order $\Lambda _4/\Lambda _5=\ell .$ Here 
$\Lambda _5={m_5}^3/\ell ^2$ is the negative cosmological
constant in the $AdS^5$ bulk. It follows that the 
non-gravitational energy of the critical bubble is
of order $E_{NG}=\Lambda _4\ell ^3.$ Using the result for 
five-dimensional Newtonian gravity, we obtain
$E_{G}=G_5M^2/\ell ^2={m_5}^3/\ell $
where $G_5={m_5}^{-3}.$ If the corrections due to
five-dimensional gravity may be characterized by the 
dimensionless parameter
$\alpha =(E_{G}/E_{NG}),$ this dimensionless parameter 
unfortunately is always of order unity.

The above order of magnitude calculation suggests that
five-dimensional gravity plays an important role in
determining the evolution of the bubbles. As the bubble
walls fluctuate, they emit gravitational waves. The
energy carried away by the gravity waves for particular mode 
is likely to comprise a substantial fraction of the energy 
originally in that mode. Moreover, gravity waves generated in the 
nucleation process may inject energy into a mode. 
Since each mode of the bubble of fixed total angular
momentum mixes linearly with an infinite number
of five-dimensional modes of the same total angular
momentum, these corrections necessarily lack an effective
(3+1) dimensional description. Moreover, unlike the 
case where a brane is surrounded by $AdS^5$ on both
sides, for which there is a decoupling of long wavelengths, 
in this case because of the $M^5$ or $dS^5$ on the 
bubble exterior, we expect no such (3+1) dimensional description
in the long wavelength limit. 
This is because $dS^5$ and $M^5,$ unlike $AdS^5,$ opens up rather
than pinching off as one passes outward from the bubble wall.
An understanding of the nature 
and importance of these corrections must await a full
(4+1)-dimensional calculation. 

Returning to the (3+1)-dimensional description, 
now consider how the fluctuations of the two expanding bubbles
translate into cosmological perturbations on the 
local brane. This is completely determined by energy-momentum
conservation\cite{Bucher,neronov,LangloisCL,btcoll}.
We assume that all the energy and the momentum 
of the two colliding branes are transfered onto
the local brane. The energy-momentum conservation here is essentially a
$(1+1)$-dimensional problem. For determining the matching
conditions at the vertex (which here is a three-dimensional spacelike
surface), it is possible to
ignore spacetime curvature because there is no
curvature singularity of co-dimension two at the vertex. Fig.~2 shows a 
cut-away  of the two bubbles colliding to
form the local brane. This cut-away is a (1+1)-dimensional plane 
orthogonal to the directions generated by 
the $SO(3,1)$ symmetry, or equivalently to the surface of collision. 
Let $\bar {\bf u}_L,$ $\bar {\bf u}_R,$ and  
$\bar {\bf u}_F$ be the tangent vectors of
the branes lying in this $(1+1)$-dimensional transverse plane, 
respectively. 
Stress-energy-momentum conservation requires that 
\begin{equation}
\rho _L\bar {\bf u}_L+
\rho _R\bar {\bf u}_R
=\rho _F\bar {\bf u}_F
\end{equation}
at the point of intersection, where
$\rho _L,$ $\rho _R,$  and $\rho _F$  
are the energy densities on the various branes
in their respective brane rest frames.
This may be explicitly demonstrated by surrounding the collision surface
by a small 
closed surface over which $T_{\mu \nu }$ is integrated. In the limit in
which this surface is small, we can ignore the effect of curvature on parallel transport
and the above conservation law follows. 
Because the bubble walls lack internal
structure, $\rho _L=\rho _R=\tau $
where $\tau $ is the surface tension of
the expanding bubble wall. However, rather than
being constant,
$\rho _F$ varies according to
how much radiation-matter is deposited 
on the local brane during the collision.

\begin{figure}
\begin{center}
\epsfxsize=2.5in
\epsfysize=2.5in
\leavevmode\epsfbox{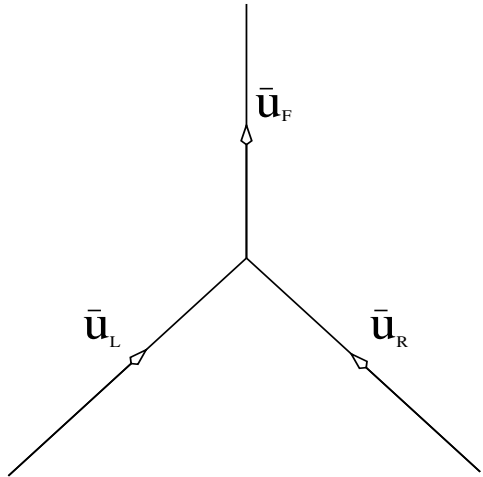}
\end{center}
\caption{{\bf Spacetime Diagram of Bubble Collision Neighborhood.}
This diagram illustrates how energy-momentum conservation determines
the outcome of the bubble collision, irrespective of the detailed microphysics
taking place at the moment of collision. Here the three transverse 
dimensions generated by the $SO(3,1)$ symmetry are orthogonal to the 
plane of the page, the vertical direction indicating time and the 
horizontal one the `fifth' spatial dimension. The 
two-dimensional vectors ${\bf \bar u}_L,$ ${\bf \bar u}_R,$ and 
${\bf \bar u}_F$ indicate the vectors tangent to the left, right,
and final branes, respectively. $\rho _L,$ $\rho _R,$ and $\rho _F,$
indicate the densities on these branes, respectively, that is the 
time-time component of the stress-energy as seen by an observer
co-moving with the brane on the plane of the page.}
\label{fig:1}
\end{figure}

Fluctuations in the fields $\chi _L$ and $\chi _R$
affect the outcome of the collision in several respects.
Firstly, they displace the point of collision with respect
to where the collision would have occurred in the absence
of perturbations. Secondly, they alter
the center-of-mass energy of the collision. Finally they
impart a transverse velocity to the local brane.
These effects are illustrated in Fig.~3.

\begin{figure}
\begin{center}
\epsfxsize=4in
\epsfysize=4in
\leavevmode\epsfbox{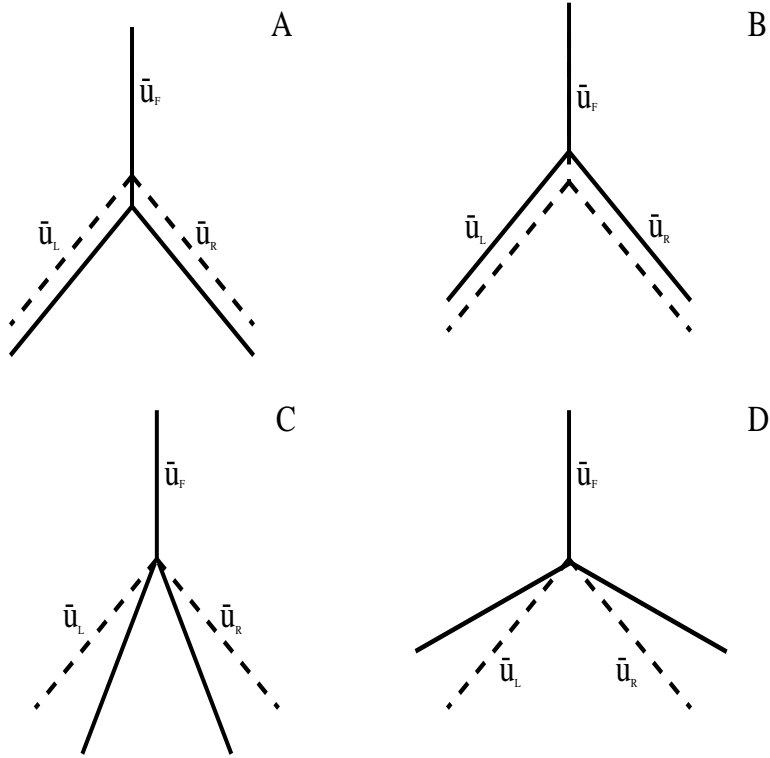}
\end{center}
\caption{{\bf Perturbations of the Bubble Collision.}
The dashed curves indicate the brane trajectories in the 
neighborhood of the collision as they would be in the absence 
of perturbations. The solid curves indicate the actual (perturbed)
brane trajectories. In panels (a) and (b) the instant of brane
collision is advanced or retarded in time. This has two effects. 
On the one hand, an earlier or later collision decreases 
or increases the velocities of the bubbles leading to an 
underdensity or overdensity at the moment of 
collision of the $(F)$ brane. On the other hand, this time delay 
warps the surface of collision, to correction to which acts in the same
direction. Because the stress-energy
content of our the local (F) brane is cooling, this time delay leads to an
underdensity or an overdensity in the respective cases. 
In (c) and (d) the spacetime position of the collision remains
unaltered but the velocities of the incident branes are slightly
decreased or increased, respectively, leading to an underdensity
or overdensity, respectively. Not shown are the possible perturbations
along the fifth dimension in position or in velocity.
Because these modes couple to the matter on the brane starting only
at quadratic order, they are not relevant to our study of linearized
perturbations. A general perturbation is a linear superposition
of all four modes.}
\label{fig:2}
\end{figure}

Before calculating the above effects quantitatively, we first describe
the degrees of freedom of the local brane. 
While the expanding bubble wall branes lack internal structure
(their stress-energy is simply $T_{\mu \nu }=\tau {g^{(4)}}_{\mu \nu }$
where $g^{(4)}$ is the four-dimensional metric induced by the
surrounding five-dimensional spacetime
and $\tau $ is the constant brane tension), the 
local brane produced in the collision possesses an 
additional degree of freedom due to the radiation-matter
deposited on the brane. This is simply the usual adiabatic
mode of the conventional FLRW cosmological models. As 
illustrated in Fig.~4, in the absence of perturbations,
the surfaces on which the universe on the local brane is at 
constant temperature are hyperboloids of constant cosmic 
time. Temperature decreases with increasing cosmic time. 
We assume that the excess energy deposited on the local brane
after the collision takes the form of the perfect fluid 
with fixed $w=p/\rho $ (taken in explicit calculations
equal to $1/3$ corresponding to a radiation-dominated universe).
In addition to this adiabatic mode, the local brane
also has a bending mode corresponding to transverse
displacements of this brane. These displacements in the 
``fifth" dimension are characterized
by a scalar field $\chi $ on the brane, just as the displacements
of the expanding bubble.
The bending mode, however, does not
affect the cosmological perturbations at linear order.\footnote{%
This bending mode is absent in the Randall-Sundrum model, constructed
using a $Z_2$ orbifold construction under which the bulk
degrees of freedom on the two sides are but redundant copies
of one another. Here, however, even though the zeroth order
solution is $Z_2$ symmetric, the degrees of freedom on the
two sides are distinct. Hence a bending mode is present.}
This is because the fields on the brane sense its bending
through the perturbation of the induced metric,
which to lowest order is quadratic.
We therefore ignore excitations of this transverse mode.

\begin{figure}
\begin{center}
\epsfxsize=2in
\epsfysize=2in
\begin{picture}(300,200)
\put(160,70){M}
\put(80,1){\leavevmode\epsfbox{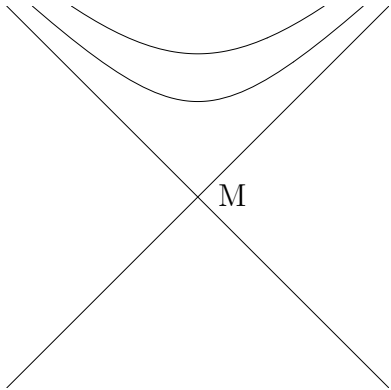}}
\end{picture}
\end{center}
\caption{{\bf Hyperplane of Local Universe (without perturbations).} 
The plane of the page corresponds to (1+1)-dimensional section of the 
(3+1)-dimensional hyperplane of points equidistant from the two nucleation 
centers.  The point $M,$ located at the vertex of the lightcone,
is the midpoint of the geodesic connecting the two 
nucleation centers. The first hyperboloid represents a section of the 
spatially hyperbolic surface of bubble collision. This surface, and 
those of constant cosmic density or temperature in its future, has a hyperbolic
spatial geometry, because it is generated by the $SO(3,1)$ residual
symmetry of the two expanding bubble geometry. The vertex acts as a
virtual ``Big-Bang" of the universe on the local brane.
Although the spatial geometry on the local brane is hyperbolic, 
parameters can be naturally chosen so that the spatial curvature radius
today lies far beyond our present horizon, rendering the universe on 
the local brane effectively flat.}
\label{fig:3}
\end{figure}

\begin{figure}
\begin{center}
\epsfxsize=3in
\epsfysize=3in
\leavevmode\epsfbox{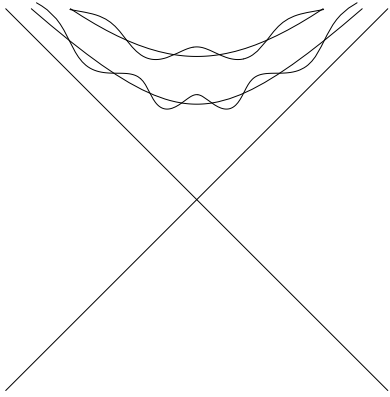}
\end{center}
\caption{{\bf Hyperplane of Local Universe (with perturbations).}
In the presence of perturbations, the surfaces of constant cosmic 
temperature become warped by the processes illustrated in Fig. 3.}
\label{fig:4}
\end{figure}

Let $\bar {\bf x}_C$ denote the two-dimensional vector
indicating the displacement of the point of collision of 
the two bubbles as a result of the perturbations
$\chi _L$ and $\chi _R,$ as indicated in Fig.~2.
Let $v_c=\tanh [2\beta ]$ be the relative velocity
of the two unperturbed bubbles before collision. 
The displacement $\bar {\bf x}_C$ is computed by
solving the two simultaneous equations 
\begin{eqnarray}
\bar {\bf x}_C&=&\chi _L\bar {\bf n}_L+\lambda _L\bar {\bf u}_L,\nonumber\\
\bar {\bf x}_C&=&\chi _R\bar {\bf n}_R+\lambda _R\bar {\bf u}_R
\label{dis}
\end{eqnarray}
where $\lambda _L$ and $\lambda _R$ are undetermined
multipliers. Here
$\bar {\bf u}_L=(\cosh \beta , \sinh \beta )$
and 
$\bar {\bf u}_R=(\cosh \beta , -\sinh \beta )$
are the vectors tangent to the respective expanding bubble
branes and 
$\bar {\bf n}_L=(\sinh \beta , \cosh \beta )$
and 
$\bar {\bf n}_R=(\sinh \beta , -\cosh \beta )$
are the respective outward normal vectors.
Solving  (\ref{dis}) yields the displacement
\begin{equation}
\bar {\bf x}_C= 
\frac{-1}{2}
\frac{(\chi _L+\chi _R)}{\sinh \beta }
\hat {\bf t} +
\frac{1}{2}
\frac{(\chi _L-\chi _R)}{\cosh \beta }
\hat {\bf e}_5.
\label{disss}
\end{equation}
The first term advances or retards the moment of 
bubble collision; the second excites the bending
mode and therefore ignored for reasons already 
discussed.

The derivatives of $\chi _L$ and $\chi _R$ alter the 
velocity of the colliding branes. Without perturbations,
the density at collision is $\rho _{coll}=2\tau \cosh [\beta ].$
To linear order, it follows that
\begin{equation}
\left(\frac{\delta \rho }{\rho }\right)_{coll}=
\frac{1}{2}\tanh [\beta ]~\delta \beta 
\end{equation}
where the perturbation in the boost parameter is
\begin{equation}
\delta \beta=\frac{\partial \chi _L}{\partial t_L}
+\frac{\partial \chi _R}{\partial t_R}-\coth [\beta ](\chi _L+\chi _R).
\end{equation}
Here $t_L$ and $t_R$ are the forward time directions normal to
the surface of collision and along the incident branes. 
The second term, calculated by setting the linear variations of the quantity
$(H_b^{-1}+\chi )\cosh \beta $ to vanish (and noting that $H_b=1$ in our 
units), reflects the fact that uniformly larger bubbles
collide earlier and consequently less energetically. 

Next we correct for the spatial warping of the surface of collision caused
by the time delay $\delta t_{pre}=-(1/\sinh \beta )(\chi _L+\chi _R)/2$
from eqn.~(\ref{disss}).
This is a time delay in a Milne universe, the spacetime prior to the bubble
collision on the plane extending the unperturbed local brane backward in time. 
To calculate $\delta \rho /\rho $ on a surface of 
constant mean spatial curvature, we apply the correction
\begin{equation}
\left(\frac{\delta \rho }{\rho }\right)_{dewarp} =
3H_{pre} (1+w)\delta t_{pre} = 
-\frac{3(1+w)}{2\sinh ^2\beta } (\chi _L+\chi _R).
\end{equation}
Here $H_{pre}$ is the expansion rate of the Milne 
universe prior to the collision.
Since $H_{pre}=1/\sinh [t]$ in units where $H_b=1,$ the above result follows. 

Combining the two contributions, we obtain\footnote{In the original version of this paper, this 
equation read
$$\left(\frac{\delta \rho }{\rho }\right)_{total}=
\frac{1}{2}\tanh [\beta ]
\left( \frac{\partial \chi _L}{\partial t_L}+
\frac{\partial \chi _R}{\partial t_R}
\right) 
-\frac{3(1+w)H_{post}}{2\sinh [\beta ]}\Biggl( \chi _L+\chi _R\Biggr) .$$
This was incorrect because $H_{post}$ should have been $H_{pre}$
and the change in the energy at collision for constant displacements
in $\chi $ had been neglected. We thank J. Garriga and T. Tanaka
for informing us of these errors.\cite{gtnew}}
\begin{equation}
\left(\frac{\delta \rho }{\rho }\right)_{total}=
\frac{1}{2}\tanh [\beta ]\left[ \left(
\frac{\partial \chi _L}{\partial t_L}+
\frac{\partial \chi _R}{\partial t_R}\right) -\coth [\beta ](\chi _L+\chi _R)\right]
-\frac{3(1+w)}{2\sinh ^2[\beta ]}\Biggl( \chi _L+\chi _R\Biggr) .
\label{bab}
\end{equation}
This formula indicates the density perturbations on a surface of constant mean 
spatial curvature. For bubble expanding in $M^5,$ $\beta =t_{coll}.$ For
those expanding in $dS^5;$ however, $\beta $ lags behind $t_{coll},$
as shall be discussed in section V. 

\section{A Hyperbolic Description of the Bunch-Davies Vacuum}

To apply the matching formula above,
it is necessary to expand the Bunch-Davies vacuum on each colliding
bubble in terms of a mode expansion natural to the colliding bubble
geometry and to that of the universe that arises on the local brane.
Since eqn.~(\ref{bab}) contains only the linear combination
$(\chi _L+\chi _R),$ it is possible to adopt the fiction that there
is only one bubble, with the perturbations at the surface of
collision determined through the formula
\begin{equation}
\frac{\delta \rho }{\rho }=
\frac{1}{\sqrt{2 \tau}}
\left[{\tanh [\beta ]}
\left( \frac{\partial \phi }{\partial t} -\coth [\beta ]\phi \right) 
-\frac{3(1+w)\phi }{\sinh ^2[\beta ]}\right]
\label{ene-field}
\end{equation}
where the field $\phi =\tau ^{1/2}(\chi _1+\chi _2)/\sqrt{2}$
has units of energy and the customary normalization.
We have adopted units in which $H_b=1$ and for bubbles expanding
in Minkowski space $\beta =t.$
For the moment we further imagine that the bubble is past and forward
eternal, extending the classical solution to the past to include 
an initially contracting phase and ignoring the bubble collision.
Now the bubble surface has the geometry of maximally extended 
(3+1)-dimensional de Sitter space $(dS^4),$ which can be covered by a series
of hyperbolic coordinates, that make manifest the $SO(3,1)$
subgroup of the full de Sitter symmetry group $SO(4,1).$ The coordinates
divide the spacetime into five patches labeled by roman numerals,
as indicated in Fig.~6. For region I the line element is
\begin{figure}
\begin{center}
\epsfxsize=3in
\epsfysize=3in
\leavevmode\epsfbox{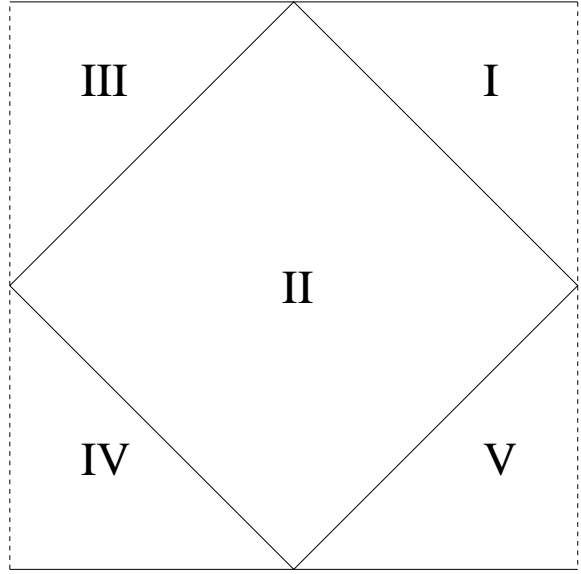}
\end{center}
\caption{{\bf Conformal Diagram of Different Hyperbolic Patches of de Sitter Space.}}
\label{fig:5}
\end{figure}
\begin{equation}
ds^2=-dt^2+\sinh ^2[t] \cdot \left[d\xi ^2+\sinh ^2[\xi ]d\Omega _{(2)}^2\right].
\end{equation}
In region II the line element is
\begin{equation}
ds^2=d\sigma ^2+\sin ^2[\sigma ] \cdot \left[-d\tau ^2+\cosh ^2[\tau ]d\Omega _{(2)}^2\right].
\end{equation}
Here we use units of length with $H_b^{-1}$ set to unity, later restoring
the correct physical units to the final results. 
With de Sitter space explicitly constructed as the embedding
$W^2+Z^2+X^2+Y^2-T^2=1$ in (4+1)-dimensional Minkowski space,
the embedding of region I is
\begin{eqnarray}
T&=&\sinh [t] \cosh [\xi ]\, , \nonumber \\
W&=&\cosh [t]\, ,\nonumber \\
Z&=&\sinh [t] \sinh [\xi ]\cos [\theta ]\, ,\nonumber \\
X&=&\sinh [t] \sinh [\xi ]\sin [\theta ]\cos [\phi ]\, , \nonumber \\
Y&=&\sinh [t] \sinh [\xi ]\sin [\theta ]\sin [\phi ].
\label{ccdd}
\end{eqnarray}
Regions III, IV, and V are essentially identical, being related 
by spatial and temporal reflections. The embedding for region II,
on the other hand, is
\begin{eqnarray}
T&=&\sin [\sigma ]\sinh [\tau ]\, ,\nonumber \\
W&=&\cos [\sigma ]\, ,\nonumber \\
Z&=&\sin [\sigma ] \cosh [\tau ]\cos [\theta ]\, ,\nonumber \\
X&=&\sin [\sigma ] \cosh [\tau ]\sin [\theta ]\cos [\phi ]\, ,\nonumber \\
Y&=&\sin [\sigma ] \cosh [\tau ]\sin [\theta ]\sin [\phi ].
\end{eqnarray}
 
Formally, many similarities are apparent between the dynamics
of the vacuum bubble expanding in (4+1) dimensions 
considered here and those of linearized quantum fluctuations of the
(3+1)-dimensional inflaton field in the single bubble open inflation\cite{open,Bucher-Turok,open-other}.
In ref.\cite{Bucher-Turok} Bucher and Turok calculated how to describe the 
Bunch-Davies
vacuum for a scalar field of arbitrary but uniform mass in terms
of a hyperbolic mode expansion. In the case of interest here, 
$m^2=-4$\cite{garrigaa,garrigaab} (in the units with $H_{b}=1$ used here).

In the sequel we first treat the $s$-wave sector of the scalar field
in isolation, using $SO(3,1)$ symmetry at the end of the calculation to generalize to all modes.
In region II, we may expand the $s$-wave sector of the scalar field 
\begin{eqnarray}
\hat \phi (\sigma , \tau )&=&
\int _{-\infty }^{+\infty }d\zeta ~F_\zeta (\sigma )
\frac{e^{-i\vert \zeta \vert \tau }}{\cosh[\tau ]}~
~\hat a_H(\zeta)
+{\rm h.c.}
\end{eqnarray}
where the spatial modes are given by
\begin{eqnarray}
F_\zeta (\sigma)=
\frac{1}{4\pi \sqrt{\vert \zeta \vert }}S(\sigma ; \zeta )
=\frac{1}{4\pi \sqrt{\vert \zeta \vert }}
\Gamma (1-i\zeta )~\frac{P^{+i\zeta }_2(\cos[\sigma])}{\sin[\sigma]} 
\end{eqnarray}
and the annihilation and creation operators satisfy the usual
commutation relations
\begin{eqnarray}
\left[~\hat a_H(\zeta),~\hat a^\dagger_H(\zeta')\right]&=&\delta(\zeta-\zeta')\,
,  \nonumber \\
\left[~\hat a_H(\zeta),~\hat a_H(\zeta')\right] &=& 0 \, , \nonumber \\
\left[~\hat a^\dagger_H(\zeta),~\hat a^\dagger_H(\zeta')\right]&=& 0 \, .
\end{eqnarray}
Here $P_2^{+i\zeta }$ is a Legendre function. The special value in the subscript
arises from the mass $m^2=-4H_b^2.$ The hyperbolic vacuum
$\vert {\rm Vac}_H\rangle ,$ a quantum state analogous to the 
well-known `Rindler' vacuum, defined by the condition that
\begin{equation}
\hat a_H(\zeta)\vert {\rm Vac }_H\rangle =0
\end{equation}
for all $\zeta $ does not coincide with the $SO(4,1)$
de Sitter invariant Bunch-Davies (BD) vacuum, $\vert {\rm Vac}_{BD}\rangle ,$
defined by the conditions
\begin{equation}
\hat a_{BD}(\zeta)\vert {\rm Vac}_{BD}\rangle =0.
\end{equation}

In ref.\cite{Bucher-Turok} it was demonstrated that the two sets of 
annihilation operators are related by the transformation
\begin{equation}
\hat a_{BD}(\zeta)
=
\frac{e^{\pi \vert \zeta \vert/2}}
{(e^{\pi \vert \zeta \vert }-e^{-\pi \vert \zeta \vert })^{1/2}}
\hat a_{H}(\zeta)
- \frac{e^{-\pi \vert \zeta \vert/2}}
{(e^{\pi \vert \zeta \vert }-e^{-\pi \vert \zeta \vert })^{1/2}}
\hat a_{H}^\dagger(\zeta) \, .
\end{equation}

Consequently, we may expand the scalar field $\hat \phi$ in terms of the
more physical BD operators as follows
\begin{eqnarray}
\hat \phi (\sigma , \tau )&=&
\int _{-\infty }^{+\infty }d\zeta ~F_\zeta (\sigma )
\frac{(e^{\pi \vert \zeta \vert/2}e^{-i\vert \zeta \vert \tau }
-e^{-\pi \vert \zeta \vert/2}e^{+i\vert \zeta \vert \tau })}
{\cosh[\tau ]~(e^{\pi \vert \zeta \vert }-e^{-\pi \vert \zeta \vert })^{1/2}}
~\hat a_{BD}(\zeta)
+{\rm h.c.}
\label{bbc} 
\end{eqnarray}

Having obtained the mode expansion of $\hat \phi $ in region II, we now
continue  $\hat \phi $ into region I, obtaining the mode expansion there in 
terms of the operators $\hat a_{BD}(\zeta)$ and their conjugates.
This is in essence a classical field theory problem, technically 
complicated by the fact that the mode functions individually diverge
near the lightcone separating regions I and II. In
terms of the variable $u$, where $\tanh[u]=\cos[\sigma],$ we note
that near this lightcone the following approximation holds
\begin{equation}
S(u; \zeta )=\frac{\Gamma (1-i\zeta )P^{+i\zeta }_{2}(\tanh [u])}{\sech [u]}
\to
\frac{e^{+i\zeta u}}{\sech [u]}\approx \sigma ^{-1}(\sigma /2)^{-i\zeta }
\label{bbb}
\end{equation}
as $u\to +\infty $ $(\sigma \to 0+)$ 
(approaching the lightcone from outside). For
this special integral value of the scalar field mass, the power
series expansion for the hypergeometric function appearing in 
the definition of the Legendre function has only a finite number
of terms, and eqn.~(\ref{bbb}) can be cast into the following
special form:
\be
S(\sigma ; \zeta )=
\frac{\Bigl(\cot [\sigma /2]\Bigr) ^{i\zeta}}{\sin [\sigma ]}\times \left[
1 - \frac{6}{(1-i\zeta)}\sin ^2[\sigma /2]
+\frac{12}{(1-i\zeta)(2-i\zeta)}\sin ^4[\sigma /2]
\right] .
\label{bbb-boo}
\ee

In ref.\cite{open} the following matching rules for the asymptotic behaviours 
from region II to region I were derived:
\begin{eqnarray}
&&\frac{e^{-i\zeta u}}{\sech [u]}
\frac{e^{+i\zeta \tau}}{\cosh [\tau ]}
\to
(+i)\frac{\sin [\zeta \xi ]}{\sinh [\xi]}e^{(+i\zeta -1)\eta }\,
, \nonumber\\
&&\frac{e^{-i\zeta u}}{\sech [u]}
\frac{e^{-i\zeta \tau}}{\cosh [\tau ]}
\to 0.
\label{bbd}
\end{eqnarray}
Two additional relations may be obtained by complex conjugation.
Here $\eta ,$ defined by $e^\eta =\tanh [t/2],$ is the region I conformal time.

In region I the temporal mode functions, which satisfy the differential 
equation
\be
\frac{d^2T}{dt^2}+3\coth [t]\frac{dT}{dt}+\left[ (-4)+\frac{(\zeta ^2+1)}
{\sinh ^2[t]}\right]T=0,
\ee
take the form
\begin{eqnarray}
T(t ; \zeta )&=&\frac{e^{\pm \pi \zeta /2}\Gamma (1-i\zeta )
P^{+i\zeta }_2(\cosh [t])}{\sinh [t]}\nonumber\\
&=&\frac{\Bigl[\coth (t/2)\Bigr] ^{i\zeta}}{\sinh t}\times \left[
1 + \frac{6}{(1-i\zeta)}\sinh ^2[t/2]
+\frac{12}{(1-i\zeta)(2-i\zeta)}\sinh ^4[t/2]
\right] .
\label{exacto}
\end{eqnarray}
The factor $e^{\pm \pi \zeta /2}$ reflects the ambiguity
of whether the Legendre function is continued above or below 
the branch point at unit argument. Our normalization is the 
geometric mean of these two continuations. 

Near the lightcone, in the $t\to 0$ or $\eta \to -\infty $ limit, 
\begin{equation}
T(t ; \zeta )\to t^{-1}(t/2)^{-i\zeta }\approx \frac{1}{2}
e^{(-i\zeta -1)\eta }.
\end{equation}
At large times, as $t\to +\infty $ (or $\eta \to 0$), 
\begin{equation}
T(t ; \zeta )\to 
\frac{(3/2)}{(2-i\zeta )(1-i\zeta)}e^t.
\label{tass}
\end{equation}

Using (\ref{bbb}), we rewrite (\ref{bbc}) in region II near the lightcone as
\begin{equation}
\hat \phi (u, \tau)= \frac{1}{4\pi}
\int _{-\infty }^{+\infty }\frac{d\zeta }{\sqrt{\vert \zeta \vert}}~
\frac{e^{+i\zeta u}}{\sech [u]}
\frac{(e^{\pi \vert \zeta \vert/2}e^{-i\vert \zeta \vert \tau }
-e^{-\pi \vert \zeta \vert/2}e^{+i\vert \zeta \vert \tau })}
{\cosh[\tau ]~(e^{\pi \vert \zeta \vert }-e^{-\pi \vert \zeta \vert })^{1/2}}
~\hat a_{BD}(\zeta)
+{\rm h.c.}\\
\label{bbe}
\end{equation}

Applying the matching rules in eqn.~(\ref{bbd}), we obtain 
in the $\eta \to -\infty $ limit of region I the expansion
\begin{eqnarray}
\hat \phi &&(\xi , \eta , \theta , \phi )= \frac{(-i)}{\sqrt{2}}
\sum_{l=0}^\infty
\sum_{m=-l }^{+l}
\int _{0}^{+\infty }\frac{d\zeta }{\sqrt{\zeta }}
\frac{1}{(e^{\pi \zeta }-e^{-\pi \zeta })^{1/2}}
R_l(\xi ;\zeta )~Y_{lm}(\theta , \phi )\nonumber\\
&&\times \Bigl[
 e^{+\pi \zeta /2}\frac{e^{(-i\zeta -1)\eta }}{2}~\hat a_{BD}(+\zeta ; l , m)
-e^{-\pi \zeta /2}\frac{e^{(+i\zeta -1)\eta }}{2}~\hat a_{BD}(-\zeta ; l , m)\Bigr]
+{\rm h.c.} \, ,
\label{bbf}
\end{eqnarray}
where we have used the $SO(3,1)$ symmetry to include the remaining non
$s$-wave modes of the spherical harmonic expansion.
Here the hyperbolic spherical functions are defined as
\be
R_l(\xi ; \zeta )=N_l(\zeta )(-)^{(l+1)}\sinh ^l[\xi ]
\frac{d^{(l+1)}}{d(\cosh \xi )^{(l+1)}}\cos [\zeta \xi ]
\ee
where $N_l(\zeta )=\Bigl[ (\pi/2)\zeta ^2(\zeta ^2+1^2)
(\zeta ^2+2^2)\ldots (\zeta ^2+l^2)\Bigr] ^{-1/2}.$ In particular,
\be
R_0(\xi ;\zeta )=\sqrt{\frac{2}{\pi}}\frac{\sin [\zeta \xi]}{\sinh [\xi ]}.
\ee
(See for example ref.~\cite{open}.)

When matched onto the exact mode functions, eqn.~(\ref{bbf}) becomes
\begin{eqnarray}
\hat \phi(\xi , &t&, \theta , \phi )=\frac{(-i)}{\sqrt{2}}
\sum_{l=0}^\infty
\sum_{m=-l}^{+l}
\int _{0}^{+\infty }\frac{d\zeta }{\sqrt{\zeta }}
\frac{1}{(e^{\pi \zeta }-e^{-\pi \zeta })^{1/2}}
R_l (\xi ;\zeta )~Y_{lm}(\theta , \phi )\nonumber\\
&&\times \Bigl[
e^{+\pi \zeta /2}T(t; \zeta )~
\hat a_{BD}(+\zeta ; l , m) 
- e^{-\pi \zeta /2}
T^*(t; \zeta )~\hat a_{BD}(-\zeta ; l , m)\Bigr]
+{\rm h.c.}
\label{bbg}
\end{eqnarray}
For a given wavenumber $\zeta ,$ the region I 
temporal dependence is as follows. 
First, for small $t,$ the mode oscillates, suffering an infinite number 
of oscillations as $t\to 0+.$ Then, after a certain time, dependent on
$\zeta $, these oscillations freeze out. In the large $t$ limit,
after the onset of this freeze out, the following asymptotic form
 becomes valid:
\begin{eqnarray}
\hat \phi &&(\xi , t, \theta , \phi )=
\frac{(-i)H_b}{\sqrt{2}}
\sum_{l=0}^\infty 
\sum_{m=-l}^{+l}
\int _{0}^{+\infty }\frac{d\zeta }{\sqrt{\zeta }}~
\frac{1}{(e^{\pi \zeta }-e^{-\pi \zeta })^{1/2}}
R_l (\xi ;\zeta )~Y_{lm}(\theta , \phi )\nonumber\\
&&\times \Bigl[
 e^{+\pi \zeta /2}\frac{(3/2)~e^t}{(2-i\zeta )(1-i\zeta )}
\hat a_{BD}(+\zeta ; l , m)
-e^{-\pi \zeta /2}\frac{(3/2)~e^t}{(2+i\zeta )(1+i\zeta )}~
\hat a_{BD}(-\zeta ; l , m)\Bigr]
+{\rm h.c..}
\label{bbh}
\end{eqnarray}
Here $H_b=(\hbar c/\ell )$ has units of energy, 
restoring $\hat \phi $ to its correct physical dimension.

\section{Scalar Power Spectrum I.: Bubbles Expanding in Minkowski Space}

Subsequent to the bubble collision we assume a radiation-matter
equation of state on our local brane. In other words, there is no inflation.
After the bubble collision, the modes, initially 
frozen in by the rapid expansion,
enter the horizon and become dynamical, first the modes of shortest wavelength
followed by those of increasingly longer wavelengths.
All modes of cosmological interest today lie far outside the horizon at the
moment of bubble collision $t=t_{bc}$. 

We expand the density contrast at the moment of bubble collision $t=t_{bc}$
\begin{equation}
\frac{\delta \rho }{\rho } (\xi , \theta , \phi )= \sqrt{2}\pi 
\sum _{l=0}^\infty \sum _{m=-l}^{+l}
\int _{0}^{+\infty }
d\zeta~{\cal A}_{lm}(\zeta)~\frac{R_l (\xi ;\zeta )}{\zeta }~
Y_{lm}(\theta , \phi )\nonumber\\
\end{equation}
where the coefficients ${\cal A}_{lm}(\zeta)$ are regarded as a 
classical Gaussian random field completely characterized by the correlators
\begin{equation}
\langle {\cal A}_{lm}(\zeta)  {\cal A}^*_{l' m'}(\zeta') \rangle = 
\frac{1}{\zeta }
\frac{d{\cal{P}}_{\delta \rho /\rho }(\zeta)}{d[\ln (\zeta)]~~}~
\delta(\zeta - \zeta')~\delta_{l,l'}~\delta_{m,m'}.
\end{equation}
Here the conventions have been chosen so that 
$d{\cal P}_{\delta \rho /\rho }/d[\ln (\zeta )]$ 
independent of $\zeta $ corresponds to a scale-free spectrum.
The terms associated with the quantum operators 
$a_{BD}(\zeta ; l , m)$ and $a_{BD}(-\zeta ; l , m)$ 
both contribute in quadrature to 
$d{\cal P}_{\delta \rho /\rho }/d[\ln (\zeta )]$. 
We apply eqn.~(\ref{ene-field}) to the temporal mode functions
given in eqn.~(\ref{exacto}), 
and squaring this in conjunction with (\ref{bbh}) gives 
\be
\frac{d{\cal P}_{\delta \rho /\rho }}{d[\ln (\zeta)]} =
\frac{1}{8\pi ^2}\cdot 
\left(\frac{H_b^4}{\tau }\right) \cdot 
\coth [\pi \zeta ] \cdot \zeta ^2 \cdot
\left|
\frac{(\zeta -2i)(\cosh [t]-i\zeta )}{(\zeta +i)\sinh [t]\cosh [t]}
-\frac{18(1+w)e^{-t}}{(2-i\zeta )(1-i\zeta )}
\right| ^2.
\label{pseqna}
\ee
For the first term enclosed within the absolute value we have used the exact
temporal dependence given in eqn.~(\ref{exacto}), whereas for the 
second dewarping term we have used the asymptotic form given in
eqn.~(\ref{tass}). For the first term the exact form was necessary,
because of a delicate cancellation. 
Because of this cancellation, pointed out by Garriga and Tanaka\cite{gtnew}
there is no `red' contribution of the form $e^{-t}/\zeta ^2,$ 
and a `blue' spectrum results. However, it should be noted that for all
scales of cosmological interest, the magnitude of this 
`blue' component is minuscule. The overall amplitude, by the dimensional
considerations discussed in sect.~II, is approximately
$H_b^4/\tau \approx (1/\Lambda _4\ell ^4)
\approx 1/(m_5\ell )^3\approx 1/(m_4\ell )^2$
and thus tiny. Even on the smallest
scales just outside the apparent horizon at the moment
of collision, with $\zeta \sim e^t,$
the fluctuations are supressed by a factor of $(m_4\ell )^{-2}.$
Therefore, the concerns of an excessive abundance of
small black holes often associated with blue spectra do not apply in
this case.  On even smaller scales, we recover the even more singular
vacuum divergences of Minkowski space associated with the 
quartic divergence of the operator $(\partial _t\phi )^2.$ These
divergences, however, do not have a cosmological significance.  

We now re-express the power spectrum above in the flat coordinates,
which are a reasonable approximation in the range
$1\ll \zeta \ll e^t.$ We set $\zeta =(k/k_c)$ where $k_c$ is the 
wavenumber of the spatial curvature. In the flat case
the expansion becomes
\begin{equation}
\frac{\delta \rho }{\rho } (r , \theta , \phi )= 
\sqrt{4\pi }\sum _{l=0}^\infty \sum _{m=-l}^{+l}
\int _{0}^{+\infty }dk~A_{l m}(k)~j_{l}(kr)~Y_{lm}(\theta , \phi )
\end{equation}
with
\begin{equation}
\langle A_{l m}(k)~A^{*}_{l m}(k') \rangle =\frac{1}{k}
\frac{dP_{\delta \rho /\rho }(k)}{d(\ln k)}~
\delta(k-k')~\delta_{l,l'}~\delta_{m,m'},
\end{equation}
the power spectrum for the density contrast becomes
\begin{equation}
\frac{dP_{\delta \rho /\rho }(k)}{d(\ln k)} =\frac{1}{2\pi^2}
\frac{{H_b}^4}{\tau }
\left( \frac{k}{k_c} \right) ^2~e^{-2t}
\cdot \left[1+9(1+w)\left( \frac{k_c}{k}\right) ^2
\right]^2 . 
\end{equation}
evaluated on superhorizon scales on a surface of unperturbed mean 
spatial curvature. In isolation the second term within the brackets 
arising from dewarping would have a red spectrum
but theis term is dominated by and interferes with the first term.
In the next section in which bubbles expanding in $dS^5$ are
discussed, we shall see that 
the cancellation in the term corresponding to the energy 
on the surface of collision persists; however, in many 
cases the dewarping term becomes dominant on large scales
and must be constrained by the CMB quadrupole. 

\section{Scalar Power Spectrum II.: Bubbles Expanding in de Sitter Space}

In the previous section we computed the scalar power spectrum for bubbles
expanding in Minkowski space $(M^5).$ In this section we generalize to the case
of a bubble expanding in de Sitter space $(dS^5)$. This calculation 
divides into two parts, one depending on the external geometry of
the collision (developed in section II for the $M^5$ case) and another depending
only on the internal geometry of the expanding bubbles (developed in 
section III). The latter part of the calculation, consisting of the 
identification
of the Bunch-Davies vacuum and its expansion in terms of an appropriate set of
modes once $m^2/H_b^2$ for the field $\chi $ has correctly been identified, 
does not depend how the $(3+1)$ dimensional
spacetime of the expanding bubble has been embedded into $(4+1)$ dimensions.
Appendix A presents an explicit demonstration that $m^2/H_b^2$ is
independent of the surrounding spacetime. On the other hand, those aspects
dealing with the external geometry are substantially altered, 
as we now explain.

\begin{figure}
\begin{center}
\epsfxsize=4in
\epsfysize=4in
\leavevmode\epsfbox{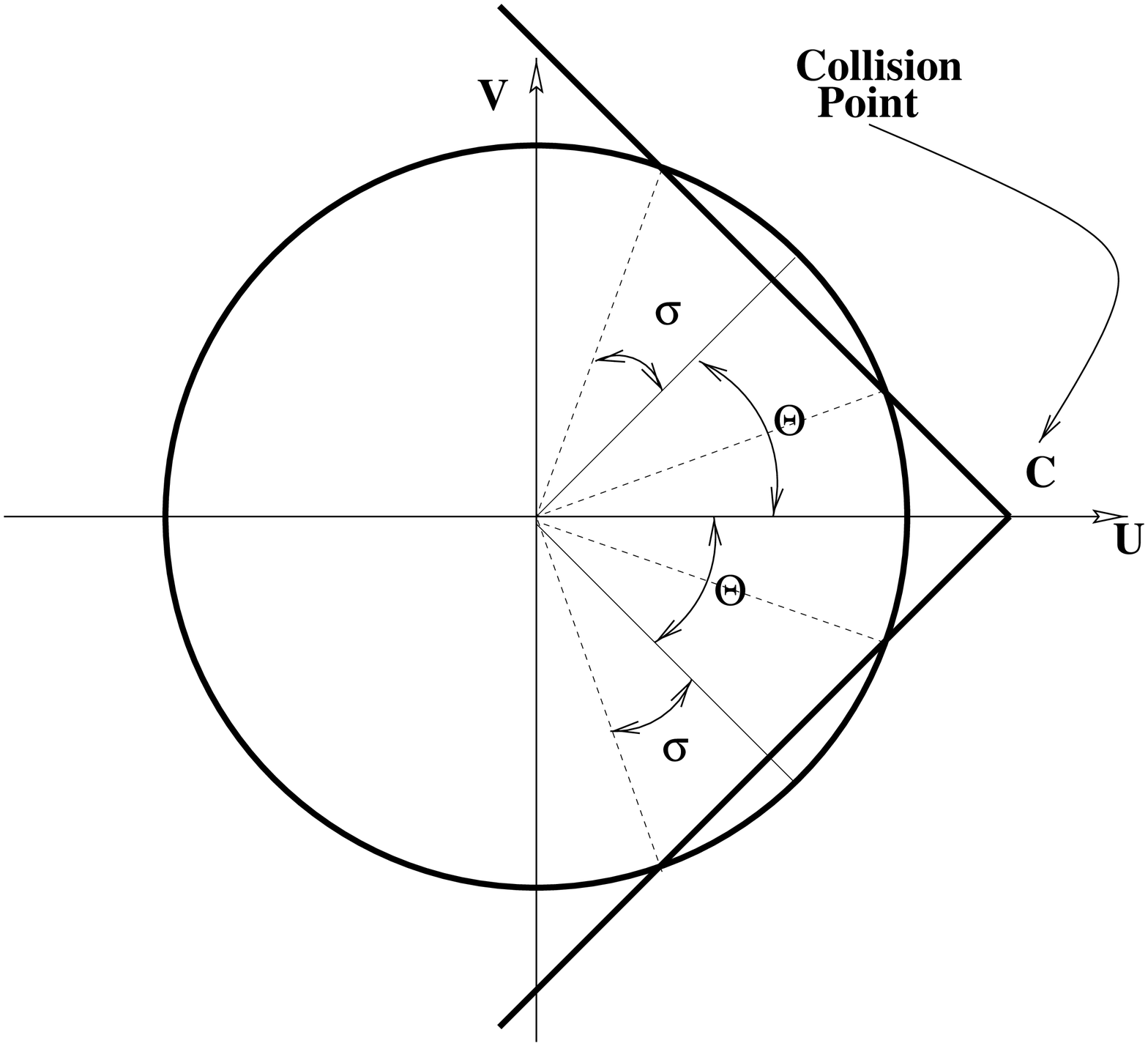}
\end{center}
\caption{{\bf
${\bf M}^6$ Representation of Two Colliding Bubbles in ${\bf dS}^5.$}
Shown in the figure is the $UV$-plane
with $T=X=Y=Z=0$ of six-dimensional Minkowski space $(M^6),$ in
which five-dimensional de Sitter space $(dS^5)$ is represented
as the invariant unit hyperboloid whose intersection with the
$UV$-plane is the unit circle shown above. The two colliding bubble
walls, with the internal geometry of $dS^4,$ are represented as
the intersection of the two planes with the unit hyperboloid
shown above as the oblique heavy lines at 
angles $\pm \Theta $ with the $V$ axis.
The invariant separation of the nucleation centers of the two bubbles
is $H_{dS^5}^{-1}\cdot (2\Theta )$ (where for convenience of calculation
the $dS^5$ geometry exterior to the bubble has been continued inside).
In reality, the portion of $dS^5$ to the right of the lines is replaced
with $AdS^5$ to represent the bubble interior. As $\Theta \to \pi /2,$ the
bubbles collide at progressively later times, eventually never colliding
when $\Theta =\pi /2$ exactly. Larger separations of the nucleation centers may be
contemplated, so that the bubbles never even come close to striking each
other in the future. This is accomplished by setting $\Theta =\pi /2$ and tilting the plane
away from the positive $T$ axis of $M^6,$ corresponding to spacelike
separation in the flat coordinate slicing of $dS^5$ so large that no
spacelike geodesic exists that connects the two nucleation centers.
The half opening angle $\sigma $ represents the radius of the critical
bubble, again in units of $H_{dS^5}^{-1}$ and adopting the fiction
of extrapolating the exterior geometry inside the bubble.
The surface of collision of the two bubbles is the represented as the
branch of the intersection of the unit hyperboloid with the line
$(U,V)=(U_c,0)=(\cos \sigma / \cos \Theta , 0)$
formed by the intersection of the two planes representing
the expanding bubbles lying in the forward time direction.
In the $UVT$-subspace
this branch consists of a single point,
but when the $XYZ$ dimensions are included, this single point
opens up into a completely spatial hyperboloid, with the
geometry of $H^3$ and a intrinsic curvature radius of 
$R=\sqrt{(\cos \Theta /\cos \sigma )^2-1}.$
The Minkowski space limit is obtained by taking
$\sigma , \Theta \to 0.$}
\label{fig:6}
\end{figure}

It is simplest to construct $dS^5$ as the
embedding of the unit hyperboloid in $M^6$ 
\begin{equation}
U^2+V^2+X^2+Y^2+Z^2-T^2=1
\end{equation}
where $ds^2=-dT^2+dU^2+dV^2+dX^2+dY^2+dZ^2$
is the usual $M^6$ metric.\cite{hawking-ellis}
The use of the redundant radial coordinate offers the technical 
advantage of rendering transformations between the 
various coordinate patches covering $dS^5$ unnecessary.
An expanding bubble with the geometry of $dS^4$ and 
an expansion rate $H_b$ in dimensionless units---that is, 
in comparison to the five-dimensional expansion
rate of the surrounding de Sitter space, which has been 
set to one---is constructed as the intersection with 
the unit hyperboloid $dS^5$ of a 
plane in $M^6$ separated from the origin of $M^6$ by an invariant
distance 
\begin{equation}
\bar U=\frac{\sqrt{H_b^2-1}}{H_b}.
\end{equation}
It is necessary that $H_b>1,$ for else there is not enough
space for the bubble to fit inside $dS^5.$

\begin{figure}
\begin{center}
\epsfxsize=3in
\epsfysize=3in
\leavevmode\epsfbox{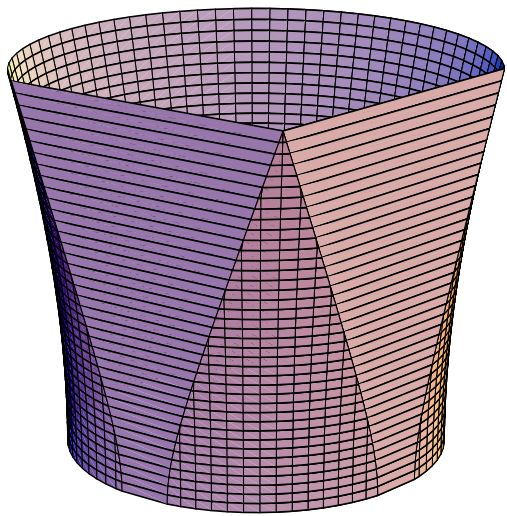}
\end{center}
\caption{{\bf Colliding Bubbles in dS${}^5$.}
Two colliding bubbles (with the geometry of $dS^4$)
expand and collide in $dS^5.$ The $dS^5$ space 
surrounding the bubbles is represented as the unit
hyperboloid in $M^6.$ The $dS^4$ expanding bubble
walls are represented as the intersection of vertical
planes with this unit hyperboloid.
}
\label{fig:8}
\end{figure}

In particular, we shall consider
collisions of bubbles nucleating at rest in the 
section defined by $T=0$ in the $M^6$ coordinates. 
These bubbles are described by the intersection of the unit hyperboloid
with planes of the form
$\cos \Theta ~U-\sin \Theta ~V=\bar V$ where 
$\Theta $ indicates the position of the 
bubble nucleation center on the unit
circle $U^2+V^2=1,$ $T=X=Y=Z=0.$ 
In terms of the usual closed hyperbolic coordinates 
for $dS^4,$ the embedding is given by 
\begin{eqnarray}
T&=&+\sqrt{1-\bar U^2}~\sinh [t],\nonumber\\ 
U&=&-\sqrt{1-\bar U^2}~\cosh [t]~\cos [\zeta ]~\sin [\Theta ]+\bar U~\cos [\Theta ],\nonumber\\ 
V&=&+\sqrt{1-\bar U^2}~\cosh [t]~\cos [\zeta ]~\cos [\Theta ]+\bar U~\sin [\Theta ],\nonumber\\ 
Z&=&+\sqrt{1-\bar U^2}~\cosh [t]~\sin [\zeta ]~\cos [\theta ],\nonumber\\ 
X&=&+\sqrt{1-\bar U^2}~\cosh [t]~\sin [\zeta ]~
\sin [\theta ]~\cos [\phi ],\nonumber\\ 
Y&=&+\sqrt{1-\bar U^2}~\cosh [t]~\sin [\zeta ]~\sin [\theta ]~
\sin [\phi ].
\label{appba}
\end{eqnarray}

We consider the collision of two such bubbles with nucleation centers
separated by an invariant geodesic distance $0<2\Theta <\pi .$\footnote{In 
computing this invariant separation we adopt the fiction that the exterior $dS^5$ geometry
extrapolates inside the bubble.}
We choose two bubbles with nucleation centers situated at the angles $+\Theta $ and 
$-\Theta $ with the positive $U$ axis.
Their surface of collision is easily calculated by noting that the 
two planes intersect on the line defined by
\begin{equation}
U=U_c=\frac{\bar U}{\cos \Theta }, \qquad  V=0
\end{equation}
where $\Theta $ must be large enough so that $\cos \Theta <\bar U$
to ensure that the bubbles do not initially overlap at $T=0.$  
The surface of collision of the two bubbles is defined by the 
$T>0$ branch of 
\begin{equation}
T^2-X^2-Y^2-Z^2=U^2-V^2-1=\frac {\bar U^2}{\cos ^2\Theta }-1=R^2>0,
\label{appbb}
\end{equation}
which is a completely spacelike hyperboloid with intrinsic curvature radius $R.$
It is useful to make the transformation $\bar U=\cos \sigma ,$ so that $\sigma $
represents the half opening angle of the bubble. It follows that
in terms of the internal bubble coordinates defined in eqn.~(\ref{appba})
\begin{equation}
T_{coll}^2=R^2=\frac{\cos ^2\sigma }{\cos ^2\Theta }-1=\sin ^2\sigma 
\sinh^2 [t_{coll}]
\end{equation}
so that 
\begin{equation}
\sinh ^2[t_{coll}]=
\frac{1}{\sin ^2\sigma }
\left[ \frac{\cos ^2\sigma -\cos ^2\Theta }{\cos ^2\Theta }\right].
\label{bbdd}
\end{equation}

To compute the relative velocity of collision we set $\zeta =\pi $ for the 
$+\Theta $ bubble and $\zeta =0$ for the $-\Theta $ bubble  
to single out the trajectories that collide at $T=R,$ $X=Y=Z=0.$
The normalized tangent vectors to these trajectories expressed in terms
of the $M^6$ coordinates are
\begin{equation}
{\cal T}_{\pm }=\cosh t \frac{~\partial }{\partial T}
+\sinh t \left (\sin \Theta \frac{~\partial }{\partial U}
                  \mp \cos \Theta \frac{~\partial }{\partial V}\right) .
\end{equation}
By taking their inner product, we determine that the relative velocity 
of the bubble walls at the collision $v_{coll},$ given by
\begin{eqnarray}
\frac{1}{\sqrt{1-v_{coll}^2}}&=&\cosh [2\beta _{coll}]\nonumber\\
&=&-{\cal T}_-\cdot {\cal T}_+=
\cosh^2[t_{coll}]+\sinh ^2[t_{coll}]\cos [2\Theta ]\nonumber\\
&=&1+2\cos ^2[\Theta ]\sinh ^2[t_{coll}].
\end{eqnarray}
The center-of-mass energy available in the collision is 
\begin{equation}
\rho _{coll}=2\tau \cosh [\beta _{coll}]=2\tau 
\sqrt{1+\cos ^2[\Theta ]\sinh ^2[t_{coll}]}
=2\tau \frac{\sin \Theta }{\sin \sigma }.
\label{appbbcc}
\end{equation}
Here $\tau $ is the density (or equivalently tension) of the expanding 
bubble brane in its rest frame. 

We compare the above result to that for Minkowski space, where
$\cos \Theta =1$ and $\rho _{coll}=2\tau \cosh [t].$ In Minkowski
space, when the expanding bubbles initially placed far enough apart, an 
arbitrarily large energy density is produced at the collision. 
In $dS^5,$ however, the most energetic possible
$\gamma $ factor is finite, attaining
$\gamma _{\infty }=\csc \sigma $ in the limit $\Theta \to \pi /2,$
in which the bubbles just barely remain within causal striking distance of each other.
In Minkowski space $\beta _{coll}$ and $\tau $ are identical, whereas in 
de Sitter space $\beta _{coll}$ falls behind $\tau $ due to, one might say,
the redshifting of the expansion of the bubble as 
a result of the expansion of the exterior $dS^5$ space. 

For bubbles expanding in de Sitter space,
it is important to keep in mind the physical distinction between
$\beta _{coll}$ and $t.$ The latter quantity is globally defined. 
It is a coordinate
of the internal geometry of the expanding $dS^4$ bubble, acting as the 
argument of the mode functions and defining relative spatial relations on this
surface. $\beta _{coll},$ by contrast, is a locally defined quantity. It is a 
property of the bubble collision, expressing the relative orientations
of the colliding bubbles. In the neighborhood of the bubble collision, for 
calculating what takes place it is justified to ignore the curvature of
$dS^5$ because displacements are small compared to $H_{dS^5}^{-1}.$
Therefore we may locally set up an approximately Minkowskian coordinate system
in a small neighborhood of the collision. Thus we calculate
the time delay
\begin{equation}
\delta \tilde t_{pre}=-\frac{1}{\sinh [\beta _{coll}]}
\frac{(\chi _L+\chi _R)}{2}
\end{equation}
identical to the result for Minkowski space, 
given in eqn.~(\ref{disss}).

The perturbation of the energy density on the surface of collision, as 
before, consists of two parts, one proportional to $\chi $ and another
proportional to $\dot \chi .$ The coefficient of the former term
is computed by taking the derivative of the logarithm of eqn.~(\ref{appbbcc}) 
with respect to $\sigma ,$ which represents the radius of the bubble in units 
in which
$H_{dS^5}^{-1}=1,$ and dividing by $\sin \sigma $ to convert to the units
of $\chi $ for which $H_{dS^5}^{-1}=1,$ yielding 
\begin{equation}
\frac{\delta \rho _{coll}}{\rho }=-\cot [\sigma ]\delta \sigma =-\cos \sigma 
\left(\frac{\chi _L+\chi _R}{2}\right) .
\end{equation}
The perturbation due to the peculiar velocities of the bubbles is
\begin{equation}
\left(\frac{\delta \rho _{coll}}{\rho }\right) _{\dot \chi }=
\tanh [\beta _{coll}]\left( \frac{\dot \chi _L+\dot \chi _R}{2}\right) 
=\sqrt{1-\frac{\sin ^2\sigma }{\sin ^2\Theta }}
\left( \frac{\dot \chi _L+\dot \chi _R}{2}\right).
\end{equation}
The derivation is exactly analogous to that for Minkowski space
except that $\tanh [t]$ is replaced with $\tanh [\beta _{coll}]$
for the reasons discussed above. 

Combining the two results, we obtain
\begin{eqnarray}
\left(\frac{\delta \rho _{coll}}{\rho }\right) _{coll}&=&
\sqrt{1-\frac{\sin ^2\sigma }{\sin ^2\Theta }}\left( \frac{\dot \chi _L+\dot \chi _R}{2}\right)
-\cos \sigma \left( \frac{\chi _L+\chi _R}{2}\right) \nonumber\\ 
&=&\cos \sigma \left[
\tanh t \left( \frac{\dot \chi _L+\dot \chi _R}{2}\right)-\left( 
\frac{\chi _L+\chi _R}{2}\right)
\right] .
\label{appbbdd}
\end{eqnarray}

We now calculate the correction due to the warping of the spatial geometry
of the surface of bubble collision.
\begin{equation}
\left(\frac{\delta \rho }{\rho }\right) _{dewarp}=3(1+w)H_{pre}~\delta 
\tilde t_{pre}
=-3(1+w)\frac{H_{pre}}
{\sinh [\beta _{coll}]}\chi \sin \sigma .
\end{equation}
The $\sin \sigma $ factor converts from the natural units of $\chi ,$
in which $H_b=1,$ to those of $dS^5$ in which its expansion rate is unity.
This correction must be added to eqn.~(\ref{appbbdd}) to obtain the density 
contrast in a gauge in which the perturbation of the trace of the 
spatial curvature vanishes on surfaces of constant cosmic time. 
Here $H_{pre},$ expressed as a sort of Hubble constant, is the extrinsic
curvature of the surface of collision as an embedding in $dS^5.$ 
We compute $H_{pre}$ as follows. The portion $U>0$ of the 
slice $V=0$ may be covered by the following one-parameter family of 
spatial hyperboloids on $dS^5,$ (with the same structure as the 
region I coordinates given in eqn.~(\ref{ccdd})) 
\begin{eqnarray}
T&=&\sinh [t_h]~\cosh [\xi ],\nonumber\\
U&=&\cosh [t_h],\nonumber\\ 
Z&=&\sinh [t_h]~\sinh [\xi ]~\cos [\theta ],\nonumber\\
X&=&\sinh [t_h]~\sinh [\xi ]~\sin [\theta ]~\cos [\phi ],\nonumber\\
Y&=&\sinh [t_h]~\sinh [\xi ]~\sin [\theta ]~ \sin [\phi ],
\label{appbsza}
\end{eqnarray}
with the line element
\be
ds^2=-d{t_h}^2+\sinh ^2[t_h]\cdot 
\Bigl[ d\xi ^2+\sinh ^2[\xi ]d\Omega _{(2)}^2\Bigr] .
\ee

It follows that $H_{pre}=\coth [t_h]$ where we set
$U=\cosh [t_h]=U_c=
\cos \sigma /\cos \Theta .$ Therefore, 
\begin{equation}
H_{pre}=\frac{\cos \sigma }{\sqrt{\cos ^2\sigma-\cos ^2\Theta }}
=\frac{\sqrt{1+\sin ^2\sigma \sinh ^2t }}{\sin \sigma \sinh t }.
\end{equation}
We use the relation (from eqns.~(\ref{bbdd}) and (\ref{appbbcc}))
\begin{equation}
\sinh [\beta _{coll}]=\frac{\cos \sigma \sinh [t_{coll}]}{\sqrt{1+\sin ^2\sigma \sinh ^2t_{coll}}}
\end{equation}
to obtain 
\begin{equation}
\left(\frac{\delta \rho }{\rho }\right) _{dewarp}=-3(1+w)
\frac{(1+\sin ^2\sigma \sinh ^2t)}{\cos \sigma \sinh ^2t}\chi .
\label{fuego}
\end{equation}

Combining eqns.~(\ref{appbbdd}) and (\ref{fuego}), we obtain a total density contrast 
perturbation
\begin{equation}
\left(\frac{\delta \rho }{\rho }\right) _{total}=\cos \sigma \left[
\tanh t \left( \frac{\dot \chi _L+\dot \chi _R}{2}\right)-\left(
\frac{\chi _L+\chi _R}{2}\right)
\right] 
-3(1+w)
\frac{(1+\sin ^2\sigma \sinh ^2t )}{\cos \sigma \sinh ^2t }\chi .
\label{fuegob}
\end{equation}
expressed in a gauge with constant mean spatial curvature. 
In light of eqn.~(\ref{fuegob}), the power spectrum for bubbles expanding in
$M^5$  given in eqn.~(\ref{pseqna}) is modified to the following
expression more generally valid for bubbles expanding in $dS^5.$
\begin{eqnarray}
\frac{d{\cal P}_{\delta \rho /\rho }}{d[\ln (\zeta)]} &=&
\frac{1}{8\pi ^2}\cdot
\left(\frac{H_b^4}{\tau }\right) \cdot
\coth [\pi \zeta ] \cdot \zeta ^2 \nonumber\\
&&\times
\left|
\frac{\cos [\sigma ]
(\zeta -2i)(\cosh [t]-i\zeta )}{(\zeta +i)\sinh [t]\cosh [t])}
-\frac{18(1+w)(1+\sin ^2[\sigma ]\sinh ^2[t])e^{-t}}{(2-i\zeta )(1-i\zeta )
\cos [\sigma ]}
\right| ^2.
\label{pseqna-ds}
\end{eqnarray}
We now consider consider the various terms of eqn.~(\ref{fuegob}).
The first term, with the exception of the $\cos (\sigma )$ 
factor, which we ignore,
is identical to the Minkowski space result. Because the mode functions for
$\chi =(\chi _L+\chi _R)/2$ depend only on the internal geometry of the 
$dS^4$ expanding bubbles, these are identical to those for the bubbles 
expanding in $M^5,$ namely those given in eqn.~(\ref{exacto}),
with the same delicate cancellation resulting. 
The second term, however, differs markedly for the case of bubbles expanding
in $dS^5$ when the colliding bubbles 
are of a size comparable to or larger than
the $dS^5$ curvature radius. To make this more manifest, 
it is useful to change to the 
more physical variables $\ell ,$ $\ell _{ext},$ and $R,$ where $\ell =H_b^{-1}$
is the critical bubble radius (as discussed in sect. II comparable to the 
$AdS^5$ curvature radius), $\ell _{ext}$ is the curvature
radius of the external $dS^5$ into which the bubble expands, and $R$ is
the curvature radius of the surface of collision, which has the geometry
of $H^3.$ Since $R=\ell _{ext}~\sin [\sigma ]\sinh [t_{coll}]$ 
and $\sin [\sigma ]=(\ell /\ell _{ext}),$ we may rewrite the second term in
the form
\be
\left( \frac{\delta \rho }{\rho }\right) _{dewarp}=
-\frac{3(1+w)}{\cos [\sigma ]}\left( 
\frac{1+(R/\ell _{ext})^2}{(R/\ell )^2}\right)\chi 
\ee
When $R\ll \ell _{ext},$ the one in the numerator dominates, yielding
the $M^5$ result of the previous section. However when 
$R\gg \ell _{ext},$ the second term dominates, leading to a larger 
amplitude for the second term with a red spectrum. 
We obtain the following power spectrum due to the dewarping, which
is the dominant contribution on large scales for bubbles expanding in $dS^5$
\be
\frac{dP_{\delta \rho /\rho }(k)}{d(\ln (k))}=
\frac{O(1)}{(m_4\ell )^2}
\left[ \frac{1+(R/\ell _{ext})^2}{(R/\ell )^2}\right] ^2
\left( \frac{R}{\ell }\right) ^2
\left( \frac{k_c}{k}\right) ^2.
\label{incendioa}
\ee
Here we have used the fact that $H_b^4/\bar \tau \approx 1/(m_4\ell )^2.$
In the $M^5$ limit, this expression is minuscule and 
well approximated as
\be
\frac{dP_{\delta \rho /\rho }(k)}{d(\ln (k))}=
\frac{O(1)}{(m_4\ell )^2}
\left( \frac{\ell }{R}\right) ^2
\left( \frac{k_c}{k}\right) ^2,
\ee
reproducing the result of the previous section.
By contrast, in the extreme $dS^5$ limit (where $R\gg l_{ext}$,
or when the bubble at collision has expanded to a size
well in excess of that of the $dS^5$ horizon), 
eqn.~(\ref{incendioa}) becomes
\be 
\frac{dP_{\delta \rho /\rho }(k)}{d(\ln (k))}=
\frac{O(1)}{(m_4\ell )^2}
\left( \frac{R^2}{(\ell ^2_{ext}/\ell )^2}\right) 
\left( \frac{k_c}{k}\right) ^2. 
\ee
The second term is small unless $R$ exceeds $\bar R,$ where
$\ell : \ell _{ext} : \bar R$ forms a geometric progression.

Physically, one can easily identify why the dewarping term 
becomes large in the $dS^5$ case in comparison to the $M^5$
case for large $R.$ In the $M^5$ case, the geometry of the 
extension of the local brane to earlier times, prior to the bubble
collision, 
has the geometry of a Milne universe, with
$H_{pre}$ rapidly falling off as $R_{coll}$ increases.
In the $dS^5$ case, however, the Milne geometry is
replaced with a $dS^4$ geometry with the same curvature
length as the exterior $dS^5.$ In the notation employed in 
this section, this extension consists of the part of the
intersection of $V=0$ with the unit hyperboloid in $M^6$
below the surface of bubble collision.
Consequently, after a certain point the decay of 
$H_{pre}$ ceases, with $H_{pre}$
approaching ${\ell _{ext}}^{-1}$ rather than zero.  

\section{Concluding Remarks}

The main conclusion of this paper is that the cosmological
perturbations generated in a colliding bubble braneworld 
are small on all observable scales, except when (1) $(m_4\ell )$
is not large (i.e., when the extra ``fifth'' dimension is 
not large, or when a substantial hierarchy between
$m_4$ and $m_5$ is lacking), or when (2) the size of the bubbles at collision
is large compared to $\bar R,$ where $\ell : \ell _{ext} : \bar R$
form a geometric progression. Here $\ell $ is the radius of
the critical bubble, approximately equal to the curvature length of the 
$AdS^5$ space inside the bubble
and $\ell _{ext}$ is the curvature length
(or apparent horizon size) of the exterior $dS^5$ into
which the bubbles expand. Unfortunately, because the 
spectrum resulting from effect (2) is red, it is not possible 
to produce the observed nearly scale-invariant 
perturbations by adjusting $R$ to lie near the edge of the 
allowed parameter space. Since the predicted spectrum dominates on large
scales, the 
most stringent constraint is imposed by the observed CMB
quadrupole. One must impose the condition
\be
\frac{1}{(m_4\ell )^2} \left( \frac{\ell }{R}\right) ^2
\left[(1+\left(\frac{R}{\ell _{ext}}\right) ^2\right] ^2
\left( \frac{k_c}{k_Q}\right) ^2
\ltorder 10^{-10}
\ee
where $k_Q$ is the dominant wavenumber contributing to the CMB quadrupole.
The perturbation calculations presented here, however, are subject to 
the caveats discussed in section II concerning the 
importance of gravitational corrections, which according to
a naive order of magnitude analysis are expected to be important
for all cases of interest.

In order to create a colliding bubble braneworld universe 
consistent with our universe, it is necessary that $R$ be large enough
so that the universe today is not dominated by curvature
and thus not empty. Calculating the minimum $R$ required so 
that the universe today is sufficiently flat involves many
uncertainties, primarily due to our ignorance of the 
equation of state on the local brane 
immediately after the collision. Nevertheless, the crude
calculation presented below is less sensitive to
these uncertainties than might be expected. If
we assume that all the energy liberated in the bubble
collision is immediately converted into radiation
and if we ignore the change as the universe cools
in the effective number
of relativistic spin degrees of freedom,
it follows that $\rho _{rad} R^4$ remains constant as
the universe expands, where $R$ is the curvature 
radius of the negatively curved $H^3$ at constant
cosmic time. Consequently, we require that
\be
\rho _{coll}R_{coll}^4\ltorder \rho _{0,rad}~R_0^4
\approx \left( \frac{R_{0,min}}{\lambda _{CMB}}
\right) ^4\approx 10^{118}
\ee
where $\lambda _{CMB}\approx 1~{\rm mm}$ is the mean wavelength
of the CMB today and $R_{0,min}$ is the smallest plausible
curvature length of the universe today consistent with 
present observation. 

For bubbles expanding in $M^5,$
$\rho _{coll}\approx \Lambda _4(R/\ell )
              \approx {m_4}^2\ell ^{-2}(R/\ell ).$
For bubbles expanding in $dS^5,$ it follows from eqn.~(\ref{appbbcc}) that
\be
\rho _{coll}\approx 
\Lambda _4\frac{(R/\ell )}{\sqrt{1+(R/\ell _{ext})^2}}
\approx \frac{{m_4}^2}{\ell ^2}
\frac{(R/\ell )}{\sqrt{1+(R/\ell _{ext})^2}}.
\ee 
In the extreme de Sitter limit $(R\gg \ell _{ext})$
in which the bubbles cease to accelerate relative to each 
other, instead attaining a subluminal limiting relative
velocity, this expression becomes
\be
\rho _{coll}=\frac{{m_4}^2}{\ell ^2}\left( \frac{\ell _{ext}}{\ell }\right) . 
\ee 
For the case of bubbles expanding in $M^5,$ we require that
\be 
(m_4\ell )^2\cdot \left( \frac{R}{\ell }\right) ^5\ltorder 10^{118}.
\ee
For bubbles expanding in $dS^5,$ this becomes
\be
(m_4\ell )^2\cdot \left( \frac{R}{\ell }\right) ^5
\frac{1}{\sqrt{1+(R/\ell _{ext})^2}}\ltorder 10^{118}.
\ee

We now examine the plausibility of obtaining a pair of colliding
bubbles with $(R/\ell )$ this large in a typical two bubble
collision. The nucleation rate for bubbles expanding in $M^5$ is 
approximately
\be
\Gamma =\ell ^{-5}\exp \Bigl[ -S_E\Bigr]
       =\ell ^{-5}\exp \Bigl[ -O(1)\cdot (m_4\ell )^2\Bigr].
\ee
Here $S_E$ is the Euclidean action of the instanton.
Setting $\Gamma \langle R\rangle ^5\approx 1$
where $\langle R\rangle $ is the mean separation of the 
colliding bubbles gives
\be \frac{\langle R\rangle }{\ell }=
\exp \left[ +\frac{O(1)}{5}\cdot (m_4\ell )^2 \right] ,
\ee
which is enormous except for small $(m_4\ell )$ 
in which the extra ``fifth'' dimension is not in any sense
large (i.e., $(m_4\ell )\ltorder 16$). For bubble nucleating in
$dS^5$ corrections to this
relation appear but we do not anticipate any difficulty in
obtaining sufficiently large bubbles.

We parenthetically note that the large values of $(R/\ell )$
required, and easily obtained owing to the exponential supression
of bubble nucleation, raise the possibility of trans-Planckian
collisions and energy densities immediately after
the bubble collision. Here the term `trans-Planckian' is taken in the
five-dimensional sense. If $\rho _{coll}\gtorder {m_5}^4,$
one might worry that the five-dimensional theory taken as a point
of departure for these calculation may cease to be valid.
Too little is presently known about trans-Planckian physics
to be able to determine whether or not this poses an obstacle to the 
colliding bubble braneworld scenario in this case. It is
quite plausible that even if trans-Planckian densities are attained,
the details of the collision matter
little in determining the perturbations on the local brane. 
It is likely that the collision may be treated as a black box, 
slightly displaced in space and time as a result of
the fluctuations of the incoming branes. By calculating the 
relative positions of these black boxes, it should be possible
to predict the outcome at late times with a minimum of 
assumptions regarding what takes place within.
In any case the initial epoch is set up on the surface of bubble collision
by the colliding bubbles in a precise manner 
as opposed to resulting from some random,
acausal singularity. 

The details of the bubble collision assumed here have been highly 
idealized.  We have assumed an instantaneous junction with no extent in 
the fifth dimension at which the two colliding bubble branes join to 
form the local brane. Reality could be much more complicated, with the junction 
extended both in time and the fifth dimension, and with a portion of the 
available energy liberated in the collision escaping into the bulk. 
For example, when the branes are formed through a scalar field,
as in the case considered by Hawking et al.\cite{fvd} and with gravitational
corrections taken into account by Wu\cite{fvd}, the ultrarelativistic
branes upon colliding undergo several bounces in the ``fifth" dimension
in which the $dS^5$ phase is restored over a region of considerable
thickness before coalescence to a single domain wall would take place.
However, since our junction conditions are based 
on no more than energy-momentum 
conservation, these conditions remain equally valid when the pointlike 
junction is replaced by a black box. If there is matter ejected into the 
bulk, if it is free falling, it never falls back onto the local brane. 
To zeroth order, as a consequence of a generalized Birkhoff's theorem, 
the effect of this matter on the bulk geometry can be absorbed into a single 
real parameter. There may be some interesting interaction between 
perturbations of such matter escaping into the bulk and the 
perturbations of the matter on the brane. 

One of the interesting features of the colliding bubble braneworld scenario
is the paucity of free parameters and unknown potentials on
which the perturbations depend. The only free parameters are
the curvature scales $\ell $ and $\ell _{ext}$ of $AdS^5$ and 
$dS^5,$ respectively, to some extent the bubble wall tension
$\tau $ (although the adjustability of this parameters is quite
limited), and to some extent the equation of state after the 
bubble collision. 

\vskip 12pt
\noindent
{\bf Acknowledgements:}
JB-P was supported by the Relativity Group PPARC Rolling Grant. MB acknowledges the 
generous support of Dennis Avery. We thank C. Carvalho, J. Garriga, S. Hawking,
N. Turok, and A. Vilenkin for useful discussions. 

{\sl Note added:} Shortly after the original version of this paper appeared,
J. Garriga and T. Tanaka announced to us that they had 
been working on calculating the perturbations in the colliding 
bubble braneworld and had also obtained a scale-invariant 
spectrum but of a different magnitude. In the resulting discussions 
it became apparent that our original paper contained
a number of errors. In particular we are grateful to 
J. Garriga and T. Tanaka for pointing out to us the 
errors noted here in the footnote of section II. 
Their revised calculation has already been placed on
the archive.\cite{gtnew}

\appendix
\section{Demonstration that $m^2/H_b^2=-4$ for Bubbles Expanding in $dS^5$}

The result due to Garriga and Vilenkin \cite{garrigaa,garrigaab} 
that $m^2=-4H_b^2$ for bubbles expanding
in five-dimensional Minkowski $(M^5)$ space is shown here to apply equally well and 
without correction to the case of bubbles expanding in five-dimensional de Sitter 
space $(dS^5).$ When $M^5$ is replaced by $dS^5,$ an extra parameter
appears, the ratio $H_{dS^5}/H_b$ (equal to zero in Minkowski space), which 
at first sight may appear to offer a possible corrections to $m^2=-4H_b^2.$ 
Using the translational symmetry of $dS^5,$ we explicitly construct a nontrivial solution of the 
equations of motion for the displacement normal to the bubble represented by
the field $\chi ,$ as defined in Section II. The form of this solution when 
expressed in terms of the coordinates internal to the $dS^4$ bubble wall geometry
is independent of $H_{dS}/H_b,$ thus proving the result. 

It is most convenient to construct $dS^5$ as the embedding
\begin{equation}
U^2+V^2+X^2+Y^2+Z^2-T^2=1
\end{equation}
where $M^6$ has the usual line element $ds^2=-dT^2+dU^2+dV^2+dX^2+dY^2+dZ^2.$
The section $U=\bar U=({\rm constant}),$ where $0<\bar U<1,$ in other
words
\begin{equation}
V^2+X^2+Y^2+Z^2-T^2=(1-\bar U^2)=(H_b/H_{dS^5})^{-2}
\end{equation}
describes the geometry of an expanding bubble with the internal geometry 
of $dS^4$ embedded in $dS^5.$
For $U>\bar U,$ the geometry actually should be that of the bubble 
interior, but here we concern ourselves only with the exterior $U<\bar U$ and 
the bubble wall itself $U=\bar U.$
In terms of the closed coordinatization of the expanding bubble, we have the embedding
\begin{eqnarray}
T&=&\sqrt{1-\bar U^2}~\sinh [\tau ],\nonumber\\
U&=&\bar U,\nonumber\\
V&=&\sqrt{1-\bar U^2}~\cosh [\tau ]~\cos [\zeta ],\nonumber\\
Z&=&\sqrt{1-\bar U^2}~\cosh [\tau ]~\sin [\zeta ]~\cos [\theta ],\nonumber\\
X&=&\sqrt{1-\bar U^2}~\cosh [\tau ]~\sin [\zeta ]~ \sin [\theta ]~\cos [\phi ],\nonumber\\
Y&=&\sqrt{1-\bar U^2}~\cosh [\tau ]~\sin [\zeta ]~\sin [\theta ]~\sin [\phi ].
\end{eqnarray}
It follows that $H_{dS^5}/H_b=\sqrt{1-\bar U^2}.$

We form a bubble translation solution $\chi $ by taking the inner product
\begin{equation}
\chi =\bar {\bf N}\cdot \bar {\bf K}
\end{equation}
where $\bar {\bf K}$ is a Killing field of $dS^5$ and $\bar {\bf N}$
is the outward unit normal vector field on the bubble on the bubble wall. 
For $\bar {\bf K}$ we arbitrarily choose the rotation
\begin{equation}
\bar {\bf K}=U\frac{~\partial}{\partial V}-V\frac{~\partial}{\partial U}. 
\end{equation}
We construct $\bar {\bf N}$ by noting that the vector field $-(\partial /\partial U)$
is everywhere normal to $dS^4$ and points out of the bubble. This field,
however, is not entirely tangent 
to $dS^5.$ It is therefore necessary to apply the 
projection operator, applying the transformation 
\begin{eqnarray}
\frac{~\partial }{\partial U}&\to &(Id-\bar {\bf R}\otimes \bar {\bf R})
\left(\frac{~\partial }{\partial U}\right) =-\left(\frac{~\partial }{\partial U}-\bar U~
\bar {\bf R}\right) \nonumber\\ 
&=&-(1-\bar U^2)\frac{~\partial }{\partial U}
-\bar U\left(
 T\frac{~\partial }{\partial T}
+V\frac{~\partial }{\partial V}
+X\frac{~\partial }{\partial X}
+Y\frac{~\partial }{\partial Y}
+Z\frac{~\partial }{\partial Z}
\right)
\label{appaa}
\end{eqnarray}
where
\begin{equation}
\bar {\bf R}
=
 T\frac{~\partial }{\partial T}
+U\frac{~\partial }{\partial U}
+V\frac{~\partial }{\partial V}
+X\frac{~\partial }{\partial X}
+Y\frac{~\partial }{\partial Y}
+Z\frac{~\partial }{\partial Z}
\end{equation}
is the field in $M^6$ normal to $dS^5.$ $\bar {\bf R}$ is normalized on $dS^5.$
Normalizing this projected field in eqn.~(\ref{appaa}), we obtain 
\begin{equation}
\bar {\bf N}=-\sqrt{1-\bar U^2}\frac{~\partial }{\partial U}
-\frac{\bar U}{\sqrt{1-\bar U^2}}
\left(
 T\frac{~\partial }{\partial T}
+V\frac{~\partial }{\partial V}
+X\frac{~\partial }{\partial X}
+Y\frac{~\partial }{\partial Y}
+Z\frac{~\partial }{\partial Z}
\right)
\end{equation}
It follows that
\begin{eqnarray}
\chi =\bar {\bf N}\cdot \bar {\bf K}
&=&
\left( U\frac{~\partial}{\partial V}-V\frac{~\partial}{\partial U},
-\sqrt{1-\bar U^2}\frac{~\partial }{\partial U}
+\frac{\bar U}{\sqrt{1-\bar U^2}}
V\frac{~\partial }{\partial V}
\right) \nonumber\\ 
&=&-\frac{\bar U^3}{\sqrt{1-\bar U^2)}})V
\end{eqnarray}
corresponds to a rigid translation of the bubble in $dS^5$. In terms of
the internal coordinates
\begin{equation}
\chi =({\rm constant})\cdot  \cosh \tau \cos \zeta 
\end{equation}
which is independent of $H_{dS^5}/H_b,$ proving that $m^2/H_b^2$ is independent 
of this ratio and thus always equal to $-4.$

\end{document}